%% file: final_version.tex
\documentclass[aps,nofootinbib,showpacs,showkeys,preprintnumbers,amsmath,amssymb]{revtex4}

\input{def_JZS.tex}

\begin{document}
\title{Dark matter candidates, helicity effects and new affine gravity with torsion}

\author{David Alvarez-Castillo$^1$}
\email{alvarez@theor.jinr.ru}

\author{Diego Julio Cirilo-Lombardo $^{12}$}
\email{diego777jcl@gmail.com, diego@theor.jinr.ru}

\author{Jilberto Zamora-Saa $^{3}$}
\email{jzamorasaa@jinr.ru}
%
\affiliation{
$^{1}$ Bogoliubov Laboratory of Theoretical Physics, Joint Institute for Nuclear Research, \\ 141980 Dubna, Moscow Region, Russian Federation.\\
$^{2}$ Universidad de Buenos Aires, Consejo Nacional de Investigaciones Cientificas y
Tecnicas (CONICET), National Institute of Plasma Physics (INFIP), Facultad de
Ciencias Exactas y Naturales, Ciudad Universitaria Buenos Aires 1428, Argentina.\\
$^{3}$ Dzhelepov Laboratory of Nuclear Problems, , Joint Institute for Nuclear Research, \\ 141980 Dubna, Moscow Region, Russian Federation.}
\begin{abstract}
The cross section for neutrino helicity spin-flip obtained from a new $f(R,T)$
model of gravitation with dynamic torsion field is phenomenologically analyzed. To this end, due to the
logarithmical energy dependence of the cross section, the relation with the
axion decay constant $f_{a}$ (Peccei-Quinn parameter) is used. Consequently
the link with the phenomenological energy/mass window is found from the
astrophysical and high energy viewpoints. The highest helicity spin-flip
cross-sectional values presented in this work coincide with a recent
estimation on the axion mass computed in the framework of finite temperature extended lattice QCD and under cosmological considerations.
\end{abstract}
\maketitle

\section{ Introduction}

In this work we study the interplay of a new gravity model with dark matter candidates. The special type of affine gravity considered here features a torsion contribution that gives rise to physical effects on fundamental particles. 
First introduced in a previous paper~\cite{Cirilo-Lombardo:2013cua}, the model provides an interaction that produces a mechanism of spin-flip that is extremely important when considering massive neutrinos. It is of the form 
$\gamma^{\alpha}j\frac{1-d}{d}\gamma_{5}h_{\alpha}$, where $h_{\alpha}$ is the torsion axial
vector, $j$ is a parameter of pure geometrical nature, and $d$ is the spacetime dimension.
Torsion effecs are important in high energy precision experiments \cite{Castillo-Felisola:2014xba} and could be studied in future beam dump experiments as ''Search for Hidden Particles'' (SHIP)  \cite{Alekhin:2015byh} and accelerators as the ''International Linear Collider'' (ILC) \cite{Castillo-Felisola:2015nma}.
The aim of this letter is to extend previous works by looking at the resulting effects of this new affine gravity from the
phenomenological and theoretical viewpoints. Therefore, here we study the spin-flip cross section caused by torsion and consider its effect on heavy neutrino oscillations (NOs)
as well as other feasible astrophysical scenarios.

As is well known, the energy dependent cross section is important when
considering neutrinos detected experimentally, because the number of events
naturally depends on the energy threshold.
In our manuscript we focus on the case of neutrinos endowed with non-standard
interactions, making it evident due to the presence of the torsion as dynamical
field, in particular the relation between the dual of the torsion field as the
gradient of the axion field, namely $h_{\alpha}\sim\nabla a$.

The organization of this paper is as follows: in Section~\ref{NAGwT} we introduce the affine gravity used in this work whereas in~\ref{SCS} we recall some of
the results of ref~\cite{Cirilo-Lombardo:2013cua}. In Section~\ref{SPI}, the phenomenological implications of
the torsion field computed before~\cite{Cirilo-Lombardo:2013cua} with respect to the neutrino
oscillation are given. In sections \ref{SAM} and~\ref{STF}, the correction to the interaction
vertex produced by the torsion field is compared with the experimental values
and the bounds to the universal parameters of the model are established. In
section~\ref{s6}, the energy window (in the Peccei Quinn sense), the relation of
the masses of the interacting fields and possible scenarios are presented.
Finally in Section~\ref{s7} we summarize the obtained results.

\section{New affine gravity with torsion}
\label{NAGwT}

In this section we review our model where the axial interaction arises. It is based on a
pure affine geometrical construction where the geometrical Lagrangian of the
theory contains dynamically the generalized curvature $\mathcal{R=}\det (%
\mathcal{R}_{\ \mu }^{a})$, namely 
\begin{equation*}
L_{g}=\sqrt{det\mathcal{R}_{\ \mu }^{a}\mathcal{R}_{a\nu }}=\sqrt{detG_{\mu
\nu }},
\end{equation*}%
characterizing a higher dimensional group manifold, e.g: SU(2,2). Then, after
the breaking of the symmetry, typically from the conformal to the Lorentz
group, e.g: $SU(2,2)\rightarrow SO\left( 1,3\right) ,$ the generalized
curvature becomes
\begin{equation*}
\mathcal{R}_{\ \mu }^{a}=\lambda \left( e_{\ \mu }^{a}+f_{\ \mu }^{a}\right)
+R_{\ \mu }^{a}\qquad \left( M_{\mu }^{a}\equiv e^{a\nu }M_{\nu \mu }\right) 
\end{equation*}%
the original Lagrangian $L_{g}$ taking the following form: 
\begin{equation}
L_{g}\rightarrow 
\sqrt{Det\left[ \lambda ^{2}\left( g_{\mu \nu }+f_{\ \mu }^{a}f_{a\nu
}\right) +2\lambda R_{\left( \mu \nu \right) }+2\lambda f_{\ \mu
}^{a}R_{[a\nu ]}+R_{\ \mu }^{a}R_{a\nu }\right] },
\end{equation}
reminiscent of a
nonlinear sigma model or M-brane. Notice that $f_{\ \mu }^{a}$ , in a sharp
contrast with the tetrad field $e_{\ \mu }^{a}$, carries the symmetry $%
e_{a\mu }f_{\ \nu }^{a}=f_{\mu \nu }=-f_{\nu \mu }$. See~\cite{CiriloLombardo:2010zza,CiriloLombardo:2011zz} for more
mathematical and geometrical details of the theory. 
Consequently, the generalized Ricci tensor splits into a symmetric and
antisymmetric part, namely: 
\begin{equation*}
R_{\mu \nu }=\overset{R_{\left( \mu \nu \right) }}{\overbrace{\overset{\circ 
}{R}_{\mu \nu }-T_{\mu \rho }^{\ \ \ \alpha }T_{\alpha \nu }^{\ \ \ \rho }}}+%
\overset{R_{\left[ \mu \nu \right] }}{\overbrace{\overset{\circ }{\nabla }%
_{\alpha }T_{\mu \nu }^{\ \ \ \alpha }}}
\end{equation*}
where $\overset{\circ }{R}_{\mu \nu }$ is the general relativistic Ricci tensor constructed with the Christoffel connection, $T_{\mu \rho }^{\ \ \ \alpha }T_{\alpha \nu }^{\ \ \ \rho }$ 
is the quadratic term in the torsion field and the antisymmetric last part 
$\overset{\circ}{\nabla }_{\alpha}T_{\mu \nu }^{\ \ \ \alpha}$
is the divergence of the totally antisymmetric torsion field that introduces its dynamics in the theory. From a
theoretical point of view our theory, containing a dynamical totally
antisymmetric torsion field, is comparable to that of Kalb-Ramond in string
or superstring theory
~\cite{Green:1987sp,Green:1987mn} but in our case energy,
matter and interactions are geometrically induced.
Notice that $^{\ast }f_{\mu \nu }$ in $L_{g}$ must be proportional to the
physical electromagnetic field, namely $jF_{\mu \nu }$ where the parameter $j
$ homogenizes the units such that the combination $g_{\mu \nu }+jF_{\mu \nu }
$ has the correct sense. Here we will not go further into details but the great advantage
of this model is that it is purely geometric, being matter, energy and
interactions geometrically induced: without energy momentum tensor added by hand.

\section{Cross-section}
\label{SCS}

Torsional effects in affine gravity manifest as a string-flip mechanism. As computed in~\cite{Cirilo-Lombardo:2013cua}, the cross section for this
process reads:%
\begin{small}
\begin{align}
\label{CS}
\sigma_{\nu}^{flip}\left(  \beta\right)     =\left(  \frac{j\mu mc}{4\hbar
}\right)  \left(  \frac{\left(  1-d\right)  ^{2}}{\pi^{2}d}\right)
^{2}\frac{4 E^{2}}{\left(  E+mc^{2}\right)  ^{2}} 
   \left[  1.09416+Ln\left(  \frac{2\left(  E^{2}-m^{2}c^{4}\right)
}{q_{\min}^{2}}\right)  \left(  Ln\left(  \frac{2\left(  E^{2}-m^{2}%
c^{4}\right)  }{q_{\min}^{2}}\right)  -0.613706\right)  \right] 
\end{align}
\end{small}%
where $j$ and $\mu$  are universal model dependent parameters, carrying units of
inverse electromagnetic field and magnetic moment respectively (e.g. $\mu\equiv$ $\zeta
\mu_{B},$ $\zeta=$cons$\tan$t$)$.  The above cross section is in fact in sharp contrast with the string
theoretical and standard model cases, depending logarithmically on the energy,
even at high energies. Notice that this cross section generalizes in some sense
the computation of reference~\cite{H.Bethe}. 
As we can see, for the explicit cross-section formula~(\ref{CS}), it is important to note the following:
\begin{enumerate}
\item Under the assumption of some astrophysical implications as presented in~\cite{Gaemers:1988fp}, 
the logarithmic terms can be bounded with values between 1 and 6,
depending on screening arguments, as generally accepted. This situation of
taking the logarithmic energy dependent terms to be constant is at present
questioned from the experimental point of view due to the arguments given in the Introduction.
\item The $j$ parameter plays formally (at the cross section level) a role
similar to that of the constant $\kappa$ of the string model with torsion.
However in our approach, it is related to some physical ''absolute field" (as
$b$ in the Born-Infeld theory), giving
the maximum value that the physical fields can take into the space-time (just as the
speed of light $c$ in the relativity theory). In such a case $j$ ("the
absolute field") will be fixed to some experimental or phenomenological value.
\item The above results can straightforwardly be applied to several physical scenarios, namely astrophysical
neutrinos, dark matter, supernovae explosions, etc.
\end{enumerate}

\section{Phenomenological implications of the model}
\label{SPI}

In the original version of the standard model (SM), leptons are grouped in three families or flavors:
\begin{equation}%
\begin{pmatrix}
\nu_{\alpha}\\
\alpha\\
\end{pmatrix}
=%
\begin{pmatrix}
\nu_{e}\\
e\\
\end{pmatrix}
;%
\begin{pmatrix}
\nu_{\mu}\\
\mu\\
\end{pmatrix}
;%
\begin{pmatrix}
\nu_{\tau}\\
\tau\\
\end{pmatrix}.
\end{equation}
While the charged leptons are massive (these get their masses via Higgs
mechanism~\cite{Bhattacharyya:2009gw}) neutrinos are not. In the middle of
60s, terrestrial experiments observed a discrepancy between the number of
neutrinos predicted by solar theoretical models and measurements of the number
of neutrinos passing through the Earth; this discrepancy was called
\textquotedblright the solar neutrino problem\textquotedblright%
\ (SNP)~\cite{Bahcall:1992hn}. A natural explanation for the SNP came from the
NOs phenomenon~\cite{Pontecorvo:1957cp} which allows the flavor transmutation
during neutrino propagation into the space. Neutrino oscillations was
confirmed by experiments and have shown that neutrinos have non-zero masses~\cite{Fukuda:1998mi,Eguchi:2002dm}. The existence of massive
neutrinos opens a new window concerning  the nature of neutrinos, Dirac or
Majorana. While Dirac neutrinos preserve the lepton number, Majorana ones
violate it by two units. Furthermore, if neutrinos are Dirac particles, right
handed ones (sometimes called steriles) are essential in order to construct
the Dirac mass term $\overline{\nu}_{L}m_{\nu}\nu_{R}$~\cite{Mohapatra:1979ia}%
. On the other hand, if they are Majorana particles, the mass term will be
$\overline{\nu}_{L}m_{\nu}\nu_{L}$ and consequently, the two ways to introduce
such gauge invariant term are: via higher dimensional operators (HDOs)~\cite{Weinberg:1979sa} 
and spontaneous symmetry breaking (SSB). It is important
to note that HDOs are not renormalizable and can be understood as an
effective theory where particles with masses $M\gg M_{W}$ have been integrated out.
However, in this section we will pay our attention on the effects of Torsion
field over the flavor neutrino oscillation.

\begin{figure}[H]
\includegraphics[width=0.47\textwidth]{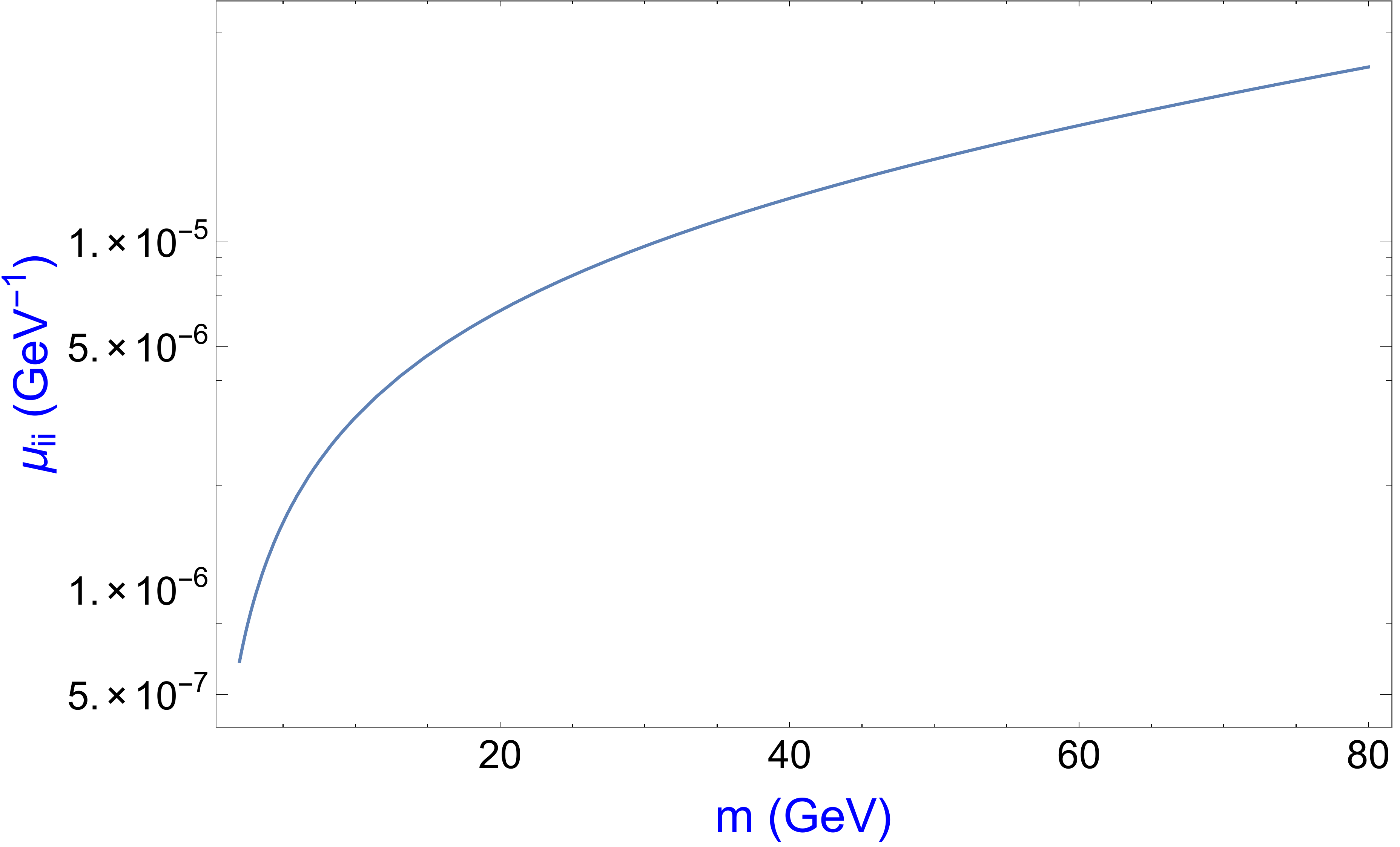} \hspace{0.1 cm}
\includegraphics[width=0.48\textwidth]{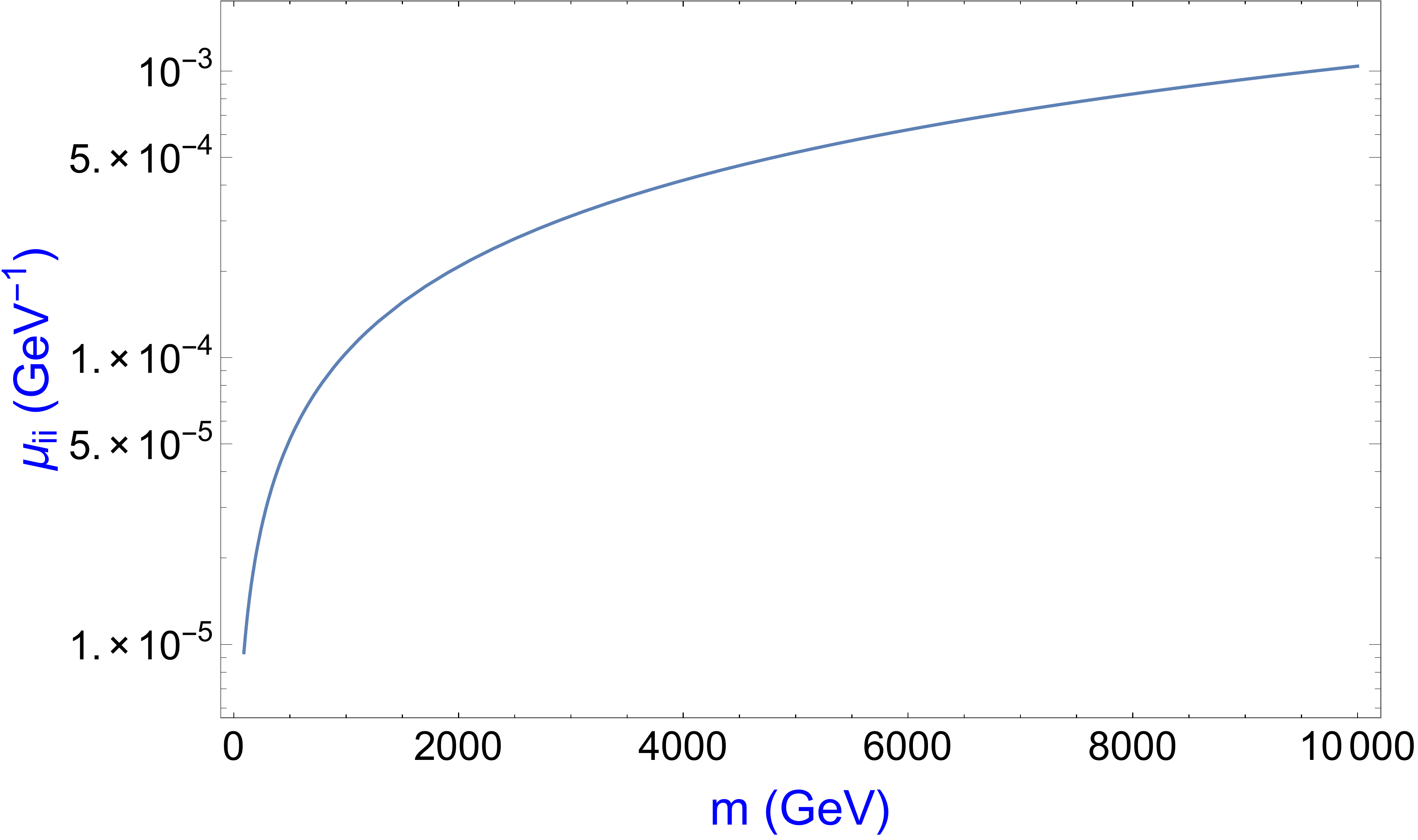}
\caption{\label{magnetic_moment}
Neutrino magnetic moment dependence on mass.
} 
\end{figure}

Let's define the flavor eigenstates, which have a defined flavor $\alpha$
\begin{equation}
|\nu_{\alpha}\rangle=\sum_{i}^{n}B_{\alpha i}|\nu_{i}\rangle,
\end{equation}
here $B_{\alpha i}$ are elements of unitary mixing matrix (PMNS-matrix) and
$|\nu_{i}\rangle$ are the mass eigenstates, which have a defined mass $m_{i}$.
The temporal evolution of the mass (or flavor) eigenstate is lead by
Schroedinger equation
\begin{equation}
\underbrace{i\frac{d}{dt}|\nu_{i}(t)\rangle=\hat{H}_{m}|\nu_{i}(t)\rangle
}_{\text{Mass Basis}}\quad;\quad\underbrace{i\frac{d}{dt}|\nu_{\alpha
}(t)\rangle=\overbrace{\hat{B}\cdot\hat{H}_{m}\cdot\hat{B}^{\dagger}}^{\hat
{H}_{f}}|\nu_{\alpha}(t)\rangle}_{\text{Flavor Basis}},
\end{equation}
where $\hat{H}_{m}$ ($\hat{H}_{f}$) is the Hamiltonian in the mass (flavor)
basis. The evolved (in time) states $|\nu_{i}(t)\rangle$ and $|\nu_{\alpha
}(t)\rangle$ are
\begin{equation}
|\nu_{i}(t)\rangle=e^{-i\hat{H}_{m}t}|\nu_{i}\rangle\quad;\quad|\nu_{\alpha
}(t)\rangle=e^{-i\hat{H}_{f}t}|\nu_{\alpha}\rangle\label{temp}.
\end{equation}
In presence of torsion-field the free Hamiltonian $\hat{H}_{0}$ will be
modified by an extra $\hat{H}_{T}$ term, as follow:
\begin{equation}
\hat{H}_{m}\ =\ \underbrace{%
\begin{pmatrix}
E_{1} & 0 & \hdots & 0\\
0 & E_{2} & \hdots & 0\\
\vdots & \vdots & \ddots & \vdots\\
0 & 0 & \hdots & E_{n}%
\end{pmatrix}
}_{\hat{H}_{0}}\quad+\quad\underbrace{%
\begin{pmatrix}
T_{11} & T_{12} & \hdots & T_{1n}\\
T_{21} & T_{22} & \hdots & T_{2n}\\
\vdots & \vdots & \ddots & \vdots\\
T_{n1} & T_{n2} & \hdots & T_{nn}%
\end{pmatrix}
}_{\hat{H}_{T}}.
\end{equation}
Due to the fact that neutrinos masses are very small ($m_{i}\ll E_{i}$), these
can be treated as relativistic particles, thus neutrino energy $E_{i}$ can be
expressed as:
\begin{equation}
E_{i}=\sqrt{m_{i}^{2}+|\vec{p}_{i}|^{2}}\approx|\vec{p}_{i}|+\frac{m_{i}^{2}%
}{2|\vec{p}_{i}|} \label{rnu}.
\end{equation}
If neutrinos are heavier (non relativistic), as in the case of heavy-sterile flavor NOs, the momenta of the two mass eigenstate are slightly different from each other, therefore the Eq.(\ref{rnu}) should be treated in a different way (see~\cite{Cvetic:2015ura} for a deepest discussion). However, since in this letter we will take care only of the phenomenological aspects, then the relativistic expression suffices. On the other hand, and in order to have a easy phenomenological discussion, we will pay attention to
scenarios with only two neutrino families ($n=2$); in such a case the rotation matrix of $SU(2)$ becomes our mixing matrix
\begin{equation}
\hat{B}=%
\begin{pmatrix}
\label{U}\cos{\theta} & \sin{\theta}\\
-\sin{\theta} & \cos{\theta}%
\end{pmatrix}.
\end{equation}
Furthermore, $\hat{H}_{m}$ can be written in terms of Pauli matrices in order
to use the group theory artillery:
\begin{align}
\hat{H}_{f}  &  \equiv\hat{B}\cdot\hat{H}_{m}\cdot\hat{B}^{\dagger}=\left(
E_{\nu}+\frac{m_{1}^{2}+m_{2}^{2}}{4E_{\nu}}+\frac{T_{11}+T_{22}}{2}\right)
\cdot\mathbb{1}_{2\times2}\label{HT}\\
&  -\tilde{\Delta}m_{D}^{2}\cdot\bigg(\sin{2\theta}\ \hat{\sigma}_{1}%
-\cos{2\theta}\ \hat{\sigma}_{3}\bigg)+T_{12}\cdot\bigg(\sin{2\theta}%
\ \hat{\sigma}_{3}+\cos{2\theta}\ \hat{\sigma}_{1}\bigg)\nonumber.
\end{align}
Here $T_{ii}\sim\frac{\mu_{ii}}{r^{2}}$, where $\mu_{ii}$ is the neutrino
magnetic moment. It is important to remark that $\mu_{ii}$ is a $2\times2$
matrix, however we will pay our attention only in the diagonal terms $\mu
_{11}$ and $\mu_{22}$ (we will assume $T_{12}=0$) in order to estimate the
impact of torsion over flavor NOs. Regarding the $\tilde{\Delta}m_{D}^{2}$
parameter, it can be understood as an \textquotedblright
effective\textquotedblright\ squared mass difference and is given as
\begin{equation}
\tilde{\Delta}m_{D}^{2}\equiv\left(  \frac{\delta m_{21}^{2}}{4E_{\nu}}%
+\frac{T_{22}-T_{11}}{2}\right)  \simeq\left(  \frac{m_{2}^{2}-m_{1}^{2}%
}{4E_{\nu}}+\frac{\mu_{22}-\mu_{11}}{2r^{2}}\right)  \label{effdelta}.
\end{equation}
It is important to note that the first term in Eq.(\ref{HT}) will not be
relevant for the NOs probabilities, due to the fact that it can only contribute
with a global phase. In order to calculate the transition probabilities we
define the flavor eigenstates ($t=0$) in matrix form as
\begin{equation}
|\nu_{\alpha}\rangle=%
\begin{pmatrix}
1\\
0
\end{pmatrix}
\quad;\quad|\nu_{\beta}\rangle=%
\begin{pmatrix}
0\\
1
\end{pmatrix}
\label{states}.
\end{equation}
By mean of Eqs.(\ref{temp},\ref{U},\ref{HT},\ref{states}) we can write the
flavor\footnote{In Eq.(\ref{testate}) we have used the Euler's formula to cast
the exponential into the binomial form.} state as
\begin{equation}
|\nu_{\alpha}(t)\rangle=\left(  \cos{\tilde{\Delta}m_{D}^{2}t}-i\sin
{\tilde{\Delta}m_{D}^{2}t}\ (\hat{\sigma}_{1}\sin{2\theta}-\hat{\sigma}%
_{3}\cos{2\theta})\right)  |\nu_{\alpha}\rangle.
\label{testate}
\end{equation}
In consequence, the probability to measure the state $|\nu_{\beta}\rangle$ at
a distance $L$ from the source, is given by
\begin{equation}
P(\nu_{\alpha}\rightarrow\nu_{\beta})=|\langle\nu_{\beta}|\nu_{\alpha
}(t=L/c)\rangle|^{2}=Sin^{2}(2\theta)\ Sin^{2}\left(  \frac{\tilde{\Delta
}m_{D}^{2}\ L}{C}\right).
\end{equation}
In order to study the energy scale, in which flavor NOs due to the torsion could play a relevant role, we should compare both terms present in Eq.~(\ref{effdelta}); it means, if both terms have the same order of magnitude the torsion effects must be taken into account in NOs, and then the energy scale becomes
\begin{equation}
\frac{\delta m_{21}^{2}}{4E_{\nu}}\sim\frac{\mu}{2r^{2}}\Rightarrow E_{\nu
}\sim\frac{\delta m_{21}^{2}r^{2}}{2\mu}. 
\label{Elcond}%
\end{equation}
In the context of Eq.~(\ref{effdelta}) we can distinguish the following cases:

\begin{itemize}
\item Reactor neutrino experiments~\cite{Agashe:2014kda} have shown that $\delta m_{21}^{2}\approx
7.53\times10^{-5}$ eV$^{2}$, whereas
$\mu\leq10^{-11}\mu_{B}\sim10^{-19}$ eV$^{-1}$ has been reported by the GEMMA expectrometer~\cite{Beda:2012zz}. 
 If we choose a $r\sim10$ km typical of supernova cores \cite{Janka:2006fh} we require
an unrealistic energy $E_{\nu}$, thus, no relevant effects of torsion are present in the supernova processes. 
\newpage
\item Most of the neutrino mass model include at least one heavy sterile neutrinos $N_i$ per leptonic family (see \cite{Mohapatra:2006gs,Mohapatra:1979ia,Cheng:1980qt,Foot:1988aq} for motivations and deeper discussions), these $N_i$ could have bigger magnetic moments which depending on the chosen model\footnote{In $SU(2)_{L}\times SU(2)_{R}\times U(1)$ (left-right symmetric models) with direct right-handed neutrino interactions (see~\cite{Czakon:1998rf,Kim:1976gk})  the massive gauge bosons states $W_1$ and $W_2$ have a dominant left-handed and right-handed coupling
\be
W_1=W_L \cos \phi -W_R \sin \phi \quad ; \quad W_2=W_L \sin \phi + W_R \cos \phi
\ee
where $\phi$ is a mixing angle and the fields $W_L$ and $W_R$ have $V \pm A$ interactions. In these models and neglecting neutrino mixing the magnetic moment $\mu$ becomes
\be
\label{LRmu}
\mu=\frac{e G_F}{2 \sqrt{2} \pi^2} \Bigg[ m_{\ell} \Bigg(1-\frac{m^2_{W_1}}{m^2_{W_2}} \Bigg)\sin \phi + \frac{3}{4} m_{\nu}\Bigg(1+\frac{m^2_{W_1}}{m^2_{W_2}} \Bigg) \Bigg].
\ee
It is important to note that term proportional to the charged lepton mass $m_{\ell}$ come from the left-right mixing and can be bigger than the second one in Eq.~(\ref{LRmu}).  On the other hand, the second term in Eq.~(\ref{LRmu}) is equivalent to the one presented in Eq.~(\ref{mu}), whose values are shown in figure~\ref{magnetic_moment}.}, could be proportional or not, to the sterile neutrino mass $m_{N_i}$ \cite{Broggini:2012df}. In the case when $\mu$ is proportional to $m_{N_i}$ the scale of energy is still very high ($\geq GUT$ scale), thus no relevant effects due to the torsion are present in NOs. However, when $\mu$ is independent of $m_{N_i}$, the scale of energy admits a fine tuning\footnote{The fine tuning $m_{N_2}-m_{N_1} \lll 1$ is fundamental in order to explain baryogenesis via leptogenesis (see \cite{Canetti:2012kh,Fukugita:1986hr,Pilaftsis:2005rv,Pilaftsis:2003gt}); in the same framework this fine tuning can be interpreted as a new symmetry in the mass Lagrangian (see \cite{Shaposhnikov:2006nn,Zamora-Saa:2016qlk}).} ($m_{N_2}-m_{N_1}\lll 1$) which can push the energy scale to lower values, in such a case the condition in Eq~(\ref{Elcond}) becomes%
\begin{equation}
E_{\nu}\sim\frac{(m_{N_2}-m_{N_1})(m_{N_2}+m_{N_1})r^{2}}{2\mu} \label{LRcond}.
\end{equation}
In scenarios of resonant CP violation~\cite{Zamora-Saa:2016qlk,Cvetic:2015naa,Cvetic:2014nla,Cvetic:2013eza,Dib:2014pga,Zamora-Saa:2016ito}, crucial for a successful theory of baryogenesis, it is found that
\be
m_{N_2}-m_{N_1} = \Gamma_N \propto |B_{\ell N}|^2 \frac{G_F^2 M_N^5} {192 \pi^3},
\label{RCP}
\ee
where $|B_{\ell N}|^2$ are the heavy-light neutrino mixings (for which it stands $|B_{\ell N}|^2 \lll 1$; the present limits are shown in \cite{Atre:2009rg}) and $G_F$ is the Fermi constant, then, the Eq.~(\ref{LRcond}) in term of Eq.~(\ref{RCP}) is
\be
E_{\nu} \sim \frac{|B_{\ell N}|^2 G_F^2 M_N^6 r^2} {192 \pi^3 \mu}
\label{lepscale}
\ee
On the other hand during the electroweak epoch\footnote{During this epoch the baryogenesis processes was started \cite{Gorbunov:2011zzc,Gorbunov:2011zz}.}, $t \sim 10^{-36}\ s $  after the big bang, the radius of the observed universe was $r \sim 10^{-2}\ m$ \cite{Ryden:2003yy}.
Then, provided that $M_N \sim 1\ GeV$, $\mu \sim 10^{-6}\ GeV^{-1}$ as it is suggest in Fig.~\ref{magnetic_moment} and $|B_{\ell N}|^2 \sim 10^{-10}$, we found an energy scale of $E_{\nu} \sim 10^{11} GeV$ which is in agreement with the energy scale of electroweak epoch presented in Fig.~6 of ref.~\cite{Lineweaver:2003ie}. Therefore, the effects of torsion in NOs could have played a significant role in the origin of baryon asymmetry of the universe via leptogenesis. However, there are extra indications (Not related with NOs) that gravity could played a role in Baryogenesis \cite{Lyth:2005jf,Alexander:2004wk,Alexander:2004xd,Alexander:2004us}.

\end{itemize}

\section{Anomalous momentum and bounds}
\label{SAM}

With the above considerations in mind, it is important to derive
the electron anomalous magnetic moment (EAM)  within this model in order to remark the role of
the torsion from the point of view of the interactions and
phenomenologically speaking. This is a key point if we want to know the
bounds over the torsion field through the physical limits over the $j$
value. Specifically, from the second order Dirac type equation (derived from
the model having into account the commutator of the full covariant
derivatives: $\nabla \sim $ $\widehat{P}_{\mu }-e\widehat{A}_{\mu
}+c_{1}\gamma ^{5}\widehat{h}_{\mu }$ ) we expand up to terms that we are interested, namely \cite{Cirilo-Lombardo:2014opa}
\begin{equation}
\sim \left\{ \left( \widehat{P}_{\mu }-e\widehat{A}_{\mu }+c_{1}\gamma ^{5}%
\widehat{h}_{\mu }\right) ^{2}-m^{2}-\frac{1}{2}\sigma ^{\mu \nu }\left[
\left( e-\omega _{1}\frac{%
\lambda }{d}\right) F_{\mu \nu }\right] \right\} u^{\lambda }=0  \tag{19}
\end{equation}%
where $\omega _{1}\frac{\lambda }{d}$ is the anomalous term. Notice that the gyromagnetic factor is modified as expected.
Although the anomalous term is clearly determined from the above equation
due the vertex correction, it is extremely useful in order to compare the
present scheme to other theoretical approaches. With these considerations in mind, it is important to derive EAM; specifically, from the last expression, one gets: $\Delta a_{e}=-%
\frac{\omega _{1}}{e}\frac{\lambda }{d}\equiv \frac{\omega _{1}}{e}\left( 1-
\frac{1}{d}\right)$. Consequently, we can see that this result is useful in
order to give constraints to the theory. The aforementioned correction  can be cast in the form
\begin{equation}
\Delta a_{e}=\left( \frac{j\mu _{B}Gmc^{4}}{4\hbar ^{2}}\right) \left( 1-
\frac{1}{d}\right) ,
\end{equation}
where we have explicitly written  the universal geometrical parameter $
\omega _{1}.$
The experimental precision measurement  of this quantity is $\Delta a_{e}^{\exp }=0.28\times 10^{-12}$ ~\cite{Peccei:1977ur}.
Therefore,  the upper bound for the universal field geometric parameter\footnote{
notice the role of the Planck length in the maximum value of $j$} $j$ is 
\be 
j < \frac{4\hbar ^{2}}{\mu _{B}Gmc^{4}}\left( \frac{d}{d-1}
\right) 0.28\times 10^{-12}.  
\ee
 Moreover, in 4-dimensions, we have as
maximum limit 
\be
 j <1.39\times 10^{-69}\ m^2  \equiv 3.4\times 10^{-56} eV^{-2} . 
\ee

\section{Torsion field and axion interaction}
\label{STF}

The particle physics phenomenology suggest that many symmetries of the nature are spontaneously broken, implying the existence of new particles, called Nambu-Goldstone bosons. Within this context, astrophysical objects, like stars or supernovae, can potentially play the role of high energy particle-physics laboratories.  In~\cite{Cirilo-Lombardo:2013cua}  we presented a \textit{concrete relation} between the axial vector $h_{\alpha}$ and the axion field $a$; furthermore, it was presented the interaction term $L_{1}\approx\frac{C_{f}}{2f_{a}}\overline{\psi}_{f}\gamma^{\alpha}\gamma_{5}\partial_{\alpha}a\psi_{f}$, where $\psi_{f}$ is a fermion field, $C_{f}$ a model-dependent
coefficient of order unity and $f_{a}$  the Peccei-Quinn energy  scale related to the
vacuum expectation value\footnote{The spontaneously broken chiral Peccei-Quinn symmetry $U_{PQ}(1)$ provides an axion field with a small mass $m_{a}=0.60\ eV\frac{10^{7}\ GeV}{f_{a}}$.} (for a more detailed discussion of the interplay between torsion and axion fields, see~\cite{Chandia:1997hu,Mercuri:2009zi,Castillo-Felisola:2015ema} and the references therein). On the other hand, within our unified gravity-model, the interaction coming from the resulting Dirac equation is:
\be
L_{int}\approx j^{\frac{1}{2}}\ \overline{\psi}_{f}\frac{1-d}{d}\gamma^{\alpha}\gamma
_{5}h_{\alpha}\psi_{f},
\ee
therefore, the above interaction is related with $L_{1}$ provided that
\begin{align}
\label{link}
\partial_{\alpha}a\sim h_{\alpha}\quad and \quad \frac{C_{f}
}{2f_{a}}\sim\frac{1-d}{d}j^{\frac{1}{2}}.
\end{align}
In addition, $C_f$ serves to define an effective Yukawa coupling of the form $g_{af}\equiv \frac{m_f C_f}{2  f_a}$. Relations in Eq.~(\ref{link}), that establish the phenomenological link between torsion and vector/axion,
shall be used for the phenomenological analysis in section \ref{s6}.

\section{Astrophysical neutrinos, oscillation and possible scenarios}
\label{s6}
\begin{figure}[H]
\includegraphics[width=0.36\textwidth]{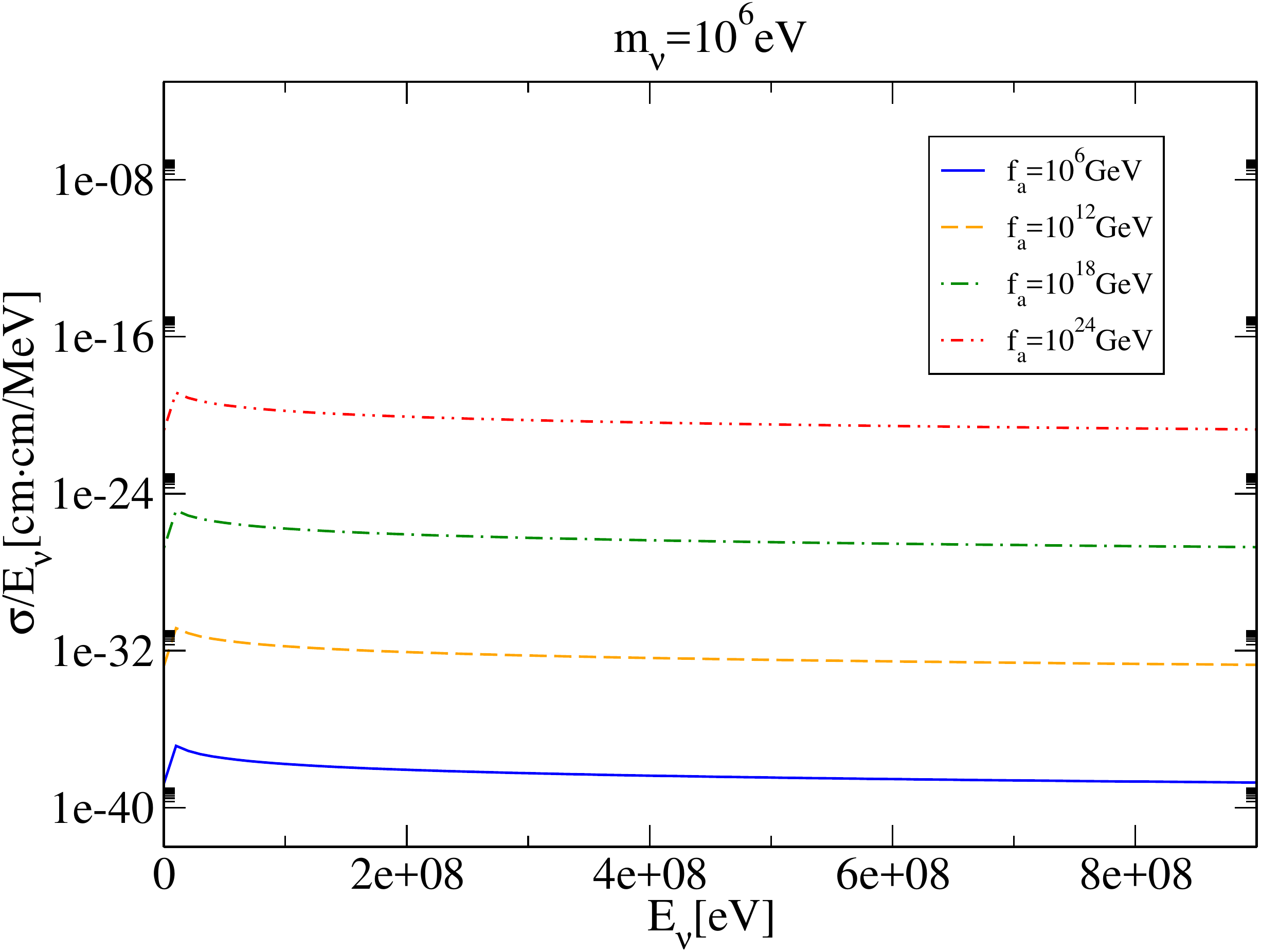} \hspace{0.5cm}
\includegraphics[width=0.36\textwidth]{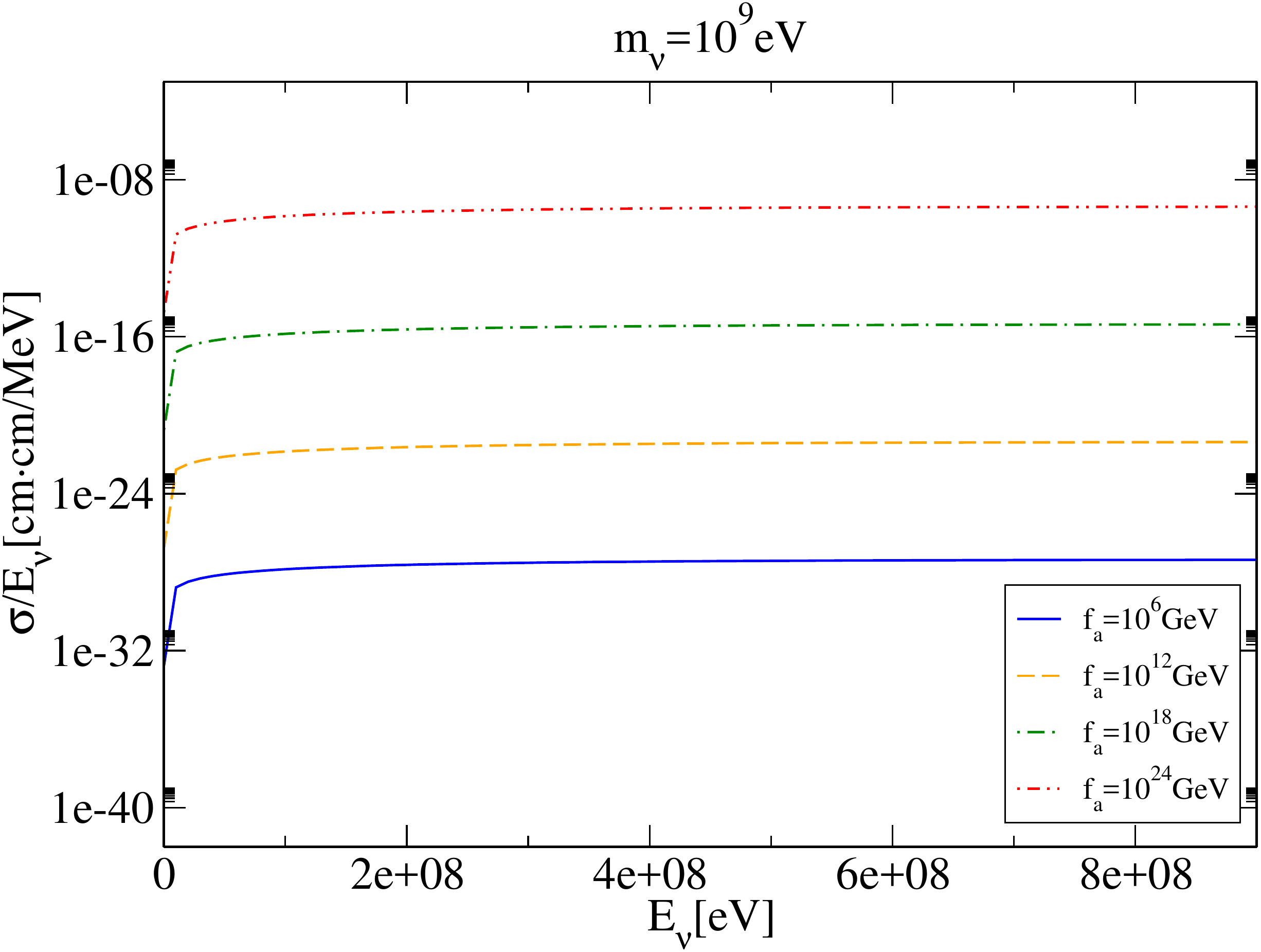}\\ \vspace{0.3cm}
\includegraphics[width=0.36\textwidth]{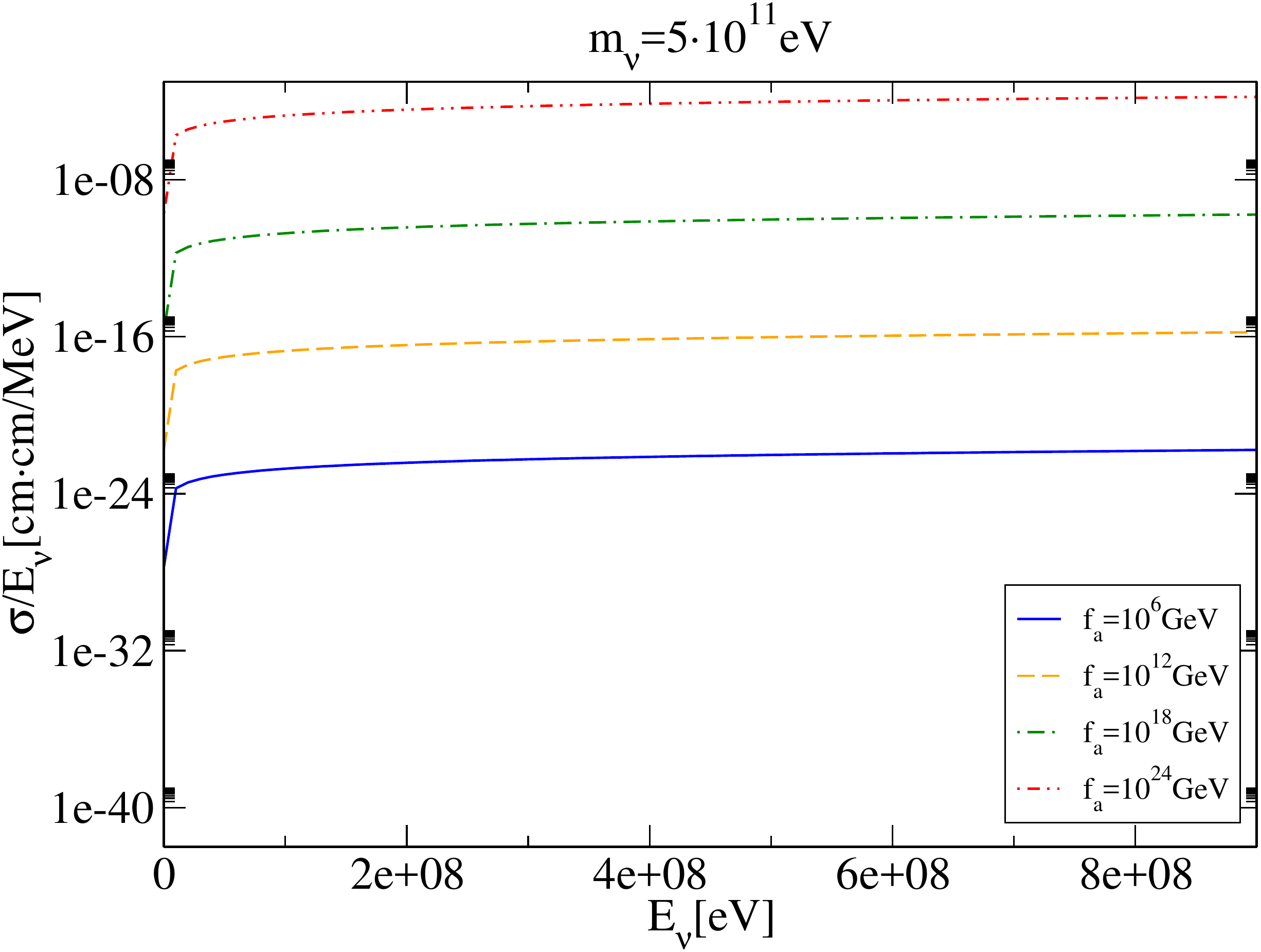}\hspace{0.5cm}
\includegraphics[width=0.36\textwidth]{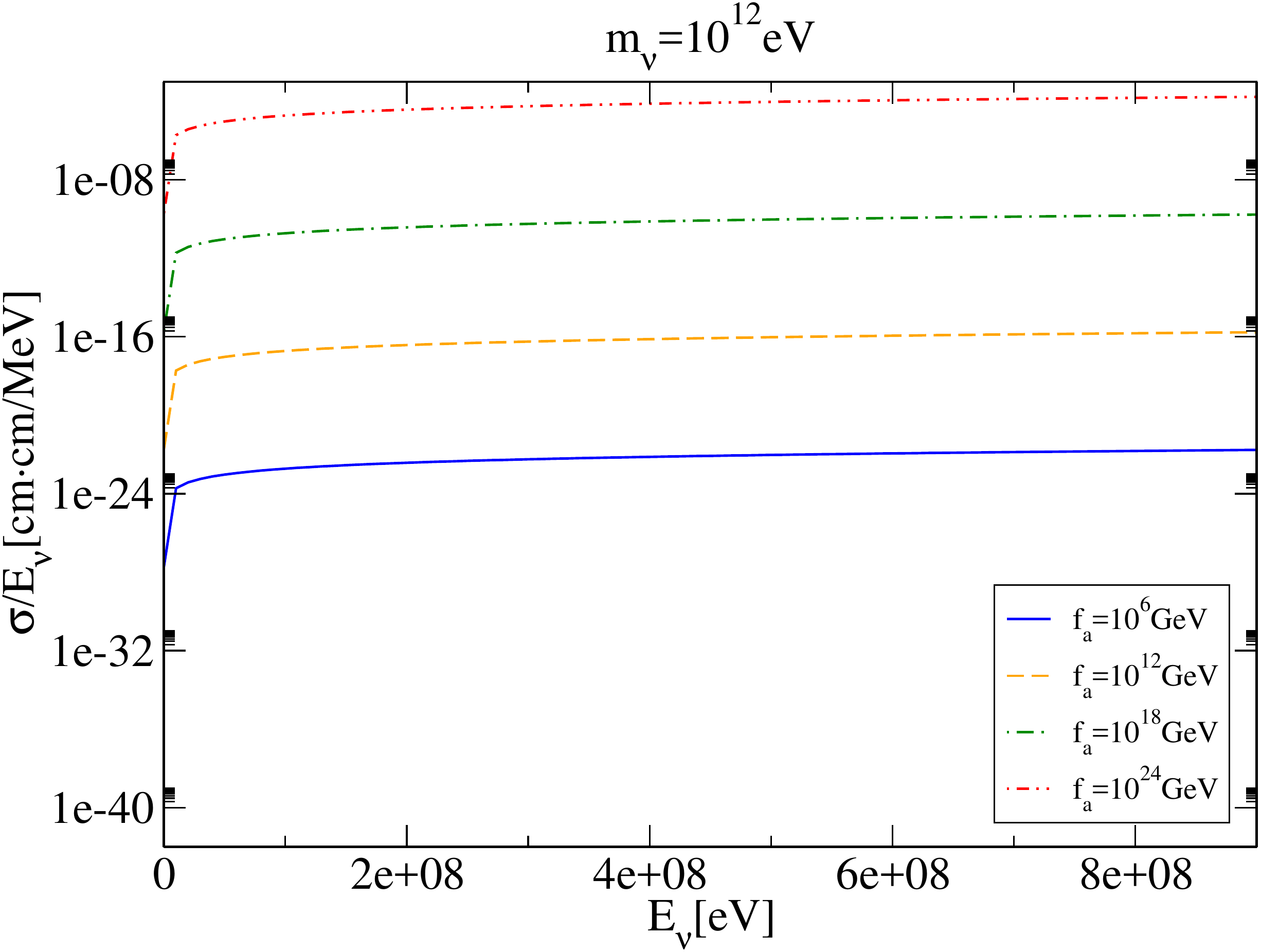}
\caption{ \textit{Set of figures for the cross section dependence on the neutrino energy featuring a fixed neutrino mass value}. 
In each plot $m_{\nu}$ is fixed whereas each line represents a given $f_{a}$. Both the mass and axion decay constant
have a direct influence on the cross section values. } 
\label{Plot_fixed_energy}
\end{figure}

\begin{figure}[H]
\includegraphics[width=0.375\textwidth]{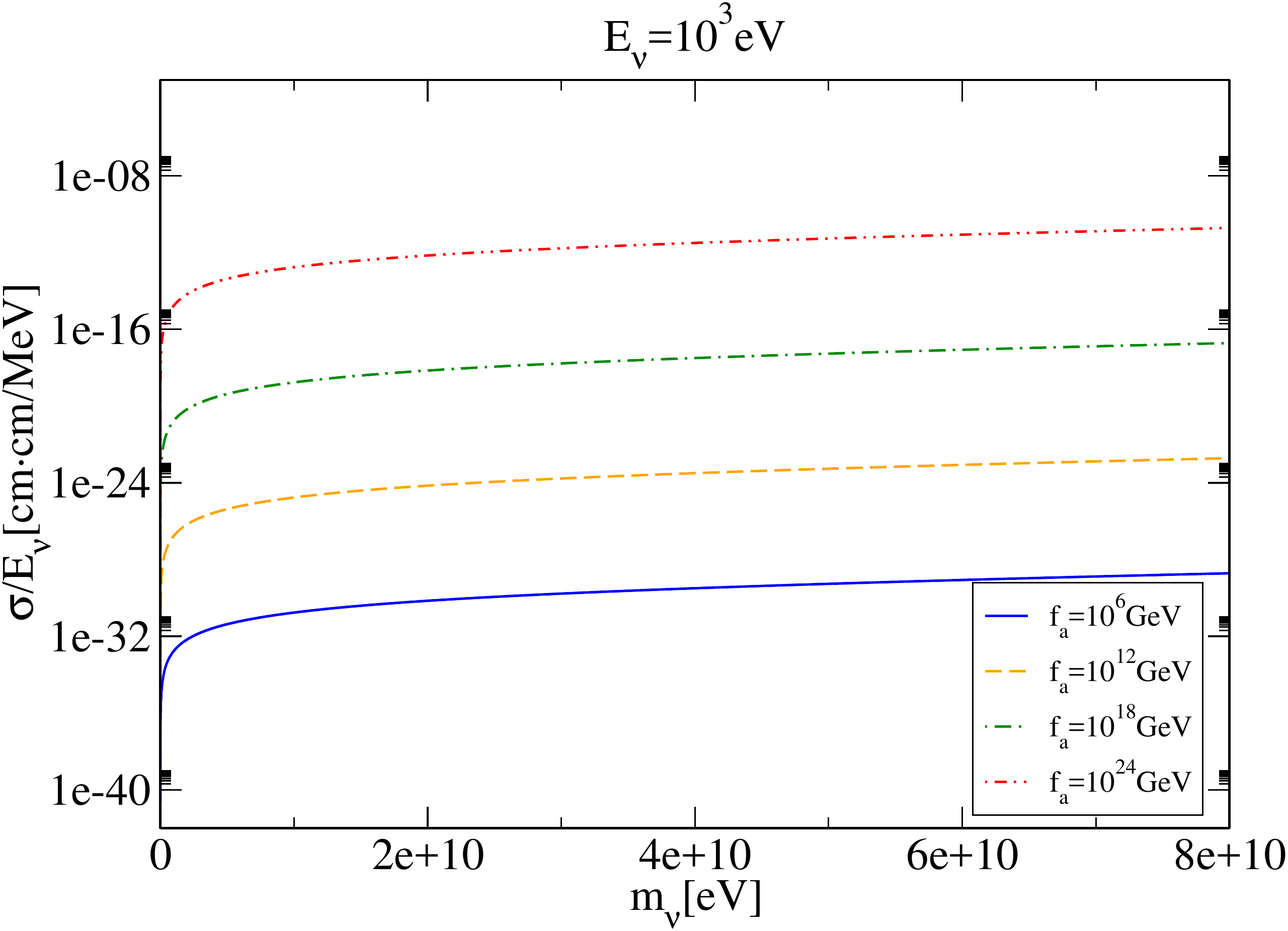} \hspace{0.5cm}
\includegraphics[width=0.375\textwidth]{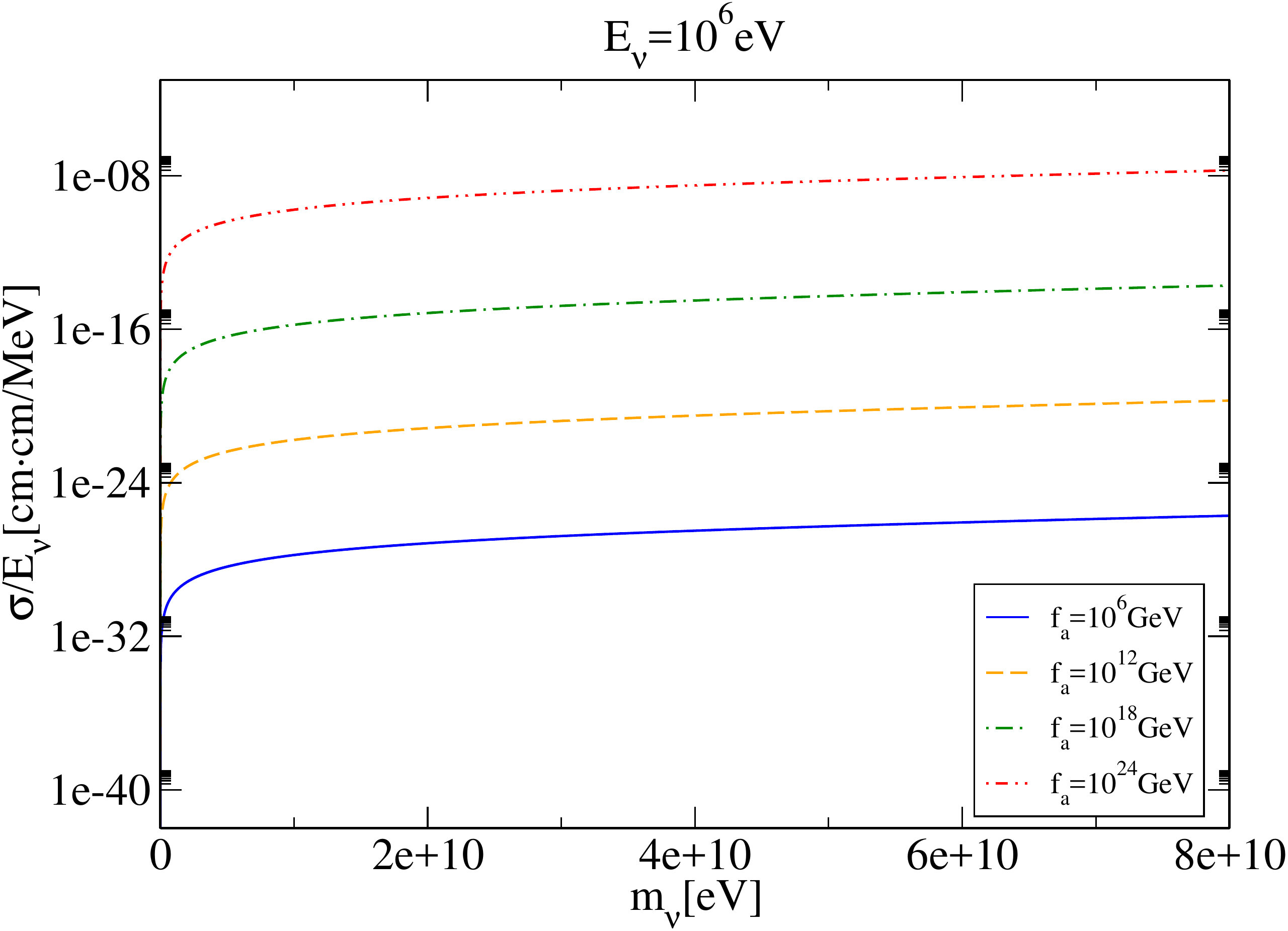}\\ \vspace{0.3cm}
\includegraphics[width=0.375\textwidth]{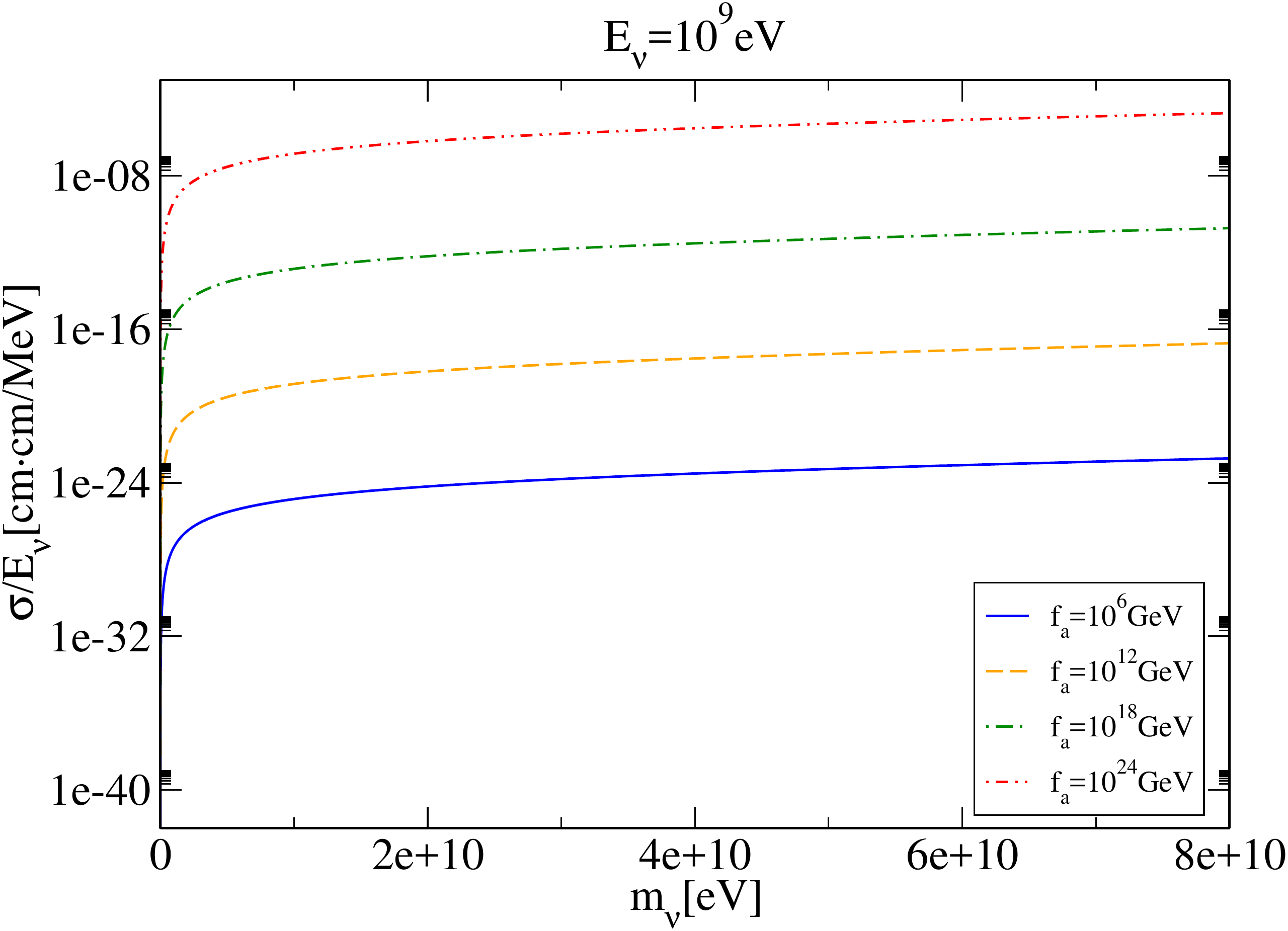} \hspace{0.5cm}
\includegraphics[width=0.375\textwidth]{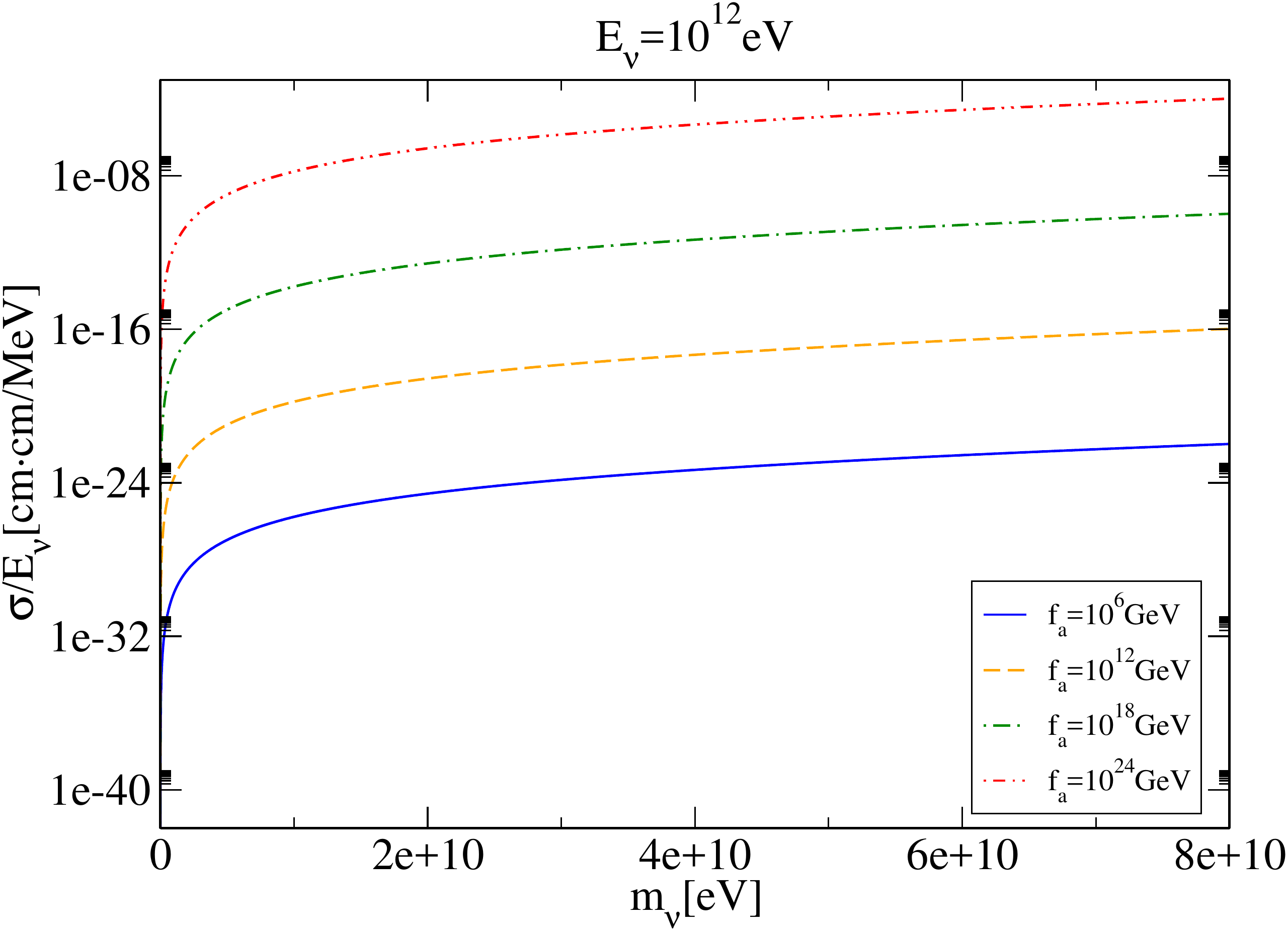}
\caption{
\textit{Set of figures for the cross section dependence on the neutrino mass featuring a fixed neutrino energy value}. 
In each plot $E_{\nu}$ is fixed whereas each line represents a given $f_{a}$. Both the energy and axion decay constant
have a direct influence on the cross section values, similar to the case of a fixed neutrino mass value.
} 
\label{Plot_fixed_mass}
\end{figure}

In order to study cross section values we look into the parameter space of
$f_{a}$, $m_{\nu}$ at different energy values. We adopt the form of the lepton
magnetic moment of a hypothetical heavy Dirac neutrino as studied
in~\cite{Dvornikov:2003js,Dvornikov:2004sj,Studenikin:2008bd}:
\begin{equation}
\mu_{\nu}={\frac{eG_{F}m_{\nu}}{8\sqrt{2}\pi^{2}}}\Bigg \{%
\begin{array}
[c]{c}%
3+{\frac{5}{6}}Q,\ m_{\ell}\ll m_{\nu}\ll M_{W},\\
1,\ \ \ \ \ \ \ \ m_{\ell}\ll M_{W}\ll m_{\nu},
\end{array}
\label{mu}
\end{equation}
where $Q=\frac{m_{\nu}^{2}}{M_{W}^{2}}$ and $M_{W}$ is the $W$ boson mass. The
axion decay constant $f_{a}$ values used in this work range from $10^{6}%
$-$10^{24}$ eV and are in agreement with~\cite{Raffelt:2006cw}, where a detailed
discussion of astrophysical and cosmological limits is presented. For
instance, in the early universe time, hot axions are expected to decouple
after the QCD epoch if $f_{a}\lesssim3\times10^{7}$ GeV; on the other hand, neutron star axion
cooling constraints suggest that $f_{a}>10^{8}$ GeV~\cite{Sedrakian:2015krq}.
Moreover, a
very interesting case is the SN 1987A pulse duration, where axion emission
might play a major role, by shortening the width of the pulsation. Energetic
emissions are characterized by values of the axion-nucleon Yukawa coupling
$g_{aN}$. Free streaming results from low $g_{aN}$ values, whereas for higher $g_{aN}$
values emission corresponds to nucleon Bremsstrahlung and is shortened until
it reaches a minimum, matching to the axion mean free path of the order
of size of the SN core. Furthermore, the highest $g_{aN}$ range will result
into axion trapping and shall be eventually emitted from the so called
\textquotedblleft axion sphere\textquotedblright. It is only after the axions
move beyond the neutrino sphere when the supernova signal becomes again
unaffected. Strongly coupled axions might interact with in-falling matter from
the supernova explosion and might lead to $\gamma$ ray emissions as
well~\cite{Raffelt:1991pw}. Under the framework of the DFSZ
model~\cite{Zhitnitsky:1980tq}, white-dwarf cooling via axion-electron
interaction is feasible for similar range of axion parameters to the supernova
case, $f_{a}\gtrsim10^{9}$ GeV. Thus, it is of great interest to explore the
impact of axion parameter values, namely the axion decay constant, to the
neutrino helicity spin-flip cross sectional values.
Figure~\ref{Plot_fixed_energy} shows the resulting cross sections as a function of the neutrino
energy $E_{\nu}$ for a set of fixed mass plots. In addition, each line
represents a chosen value of the axion decay constant $f_{a}$. The general
trend is that the larger the neutrino masses the larger the cross section
values. The cross section seems to be more dramatically dependent on the axion
decay constant, presenting the same behavior as for the mass dependence. In
figure~\ref{Plot_fixed_mass} the neutrino energy is fixed whereas the mass becomes the free
parameter. The result is the same, the $f_{a}$ parameter plays a mayor role in
the determination of cross section values. \\

Moreover, it is worth noticing that a recent estimation on the axion mass has been computed
in~\cite{Borsanyi:2016ksw} in the framework of finite temperature extended lattice QCD and
under cosmological considerations. The result is a value of the axion mass in
the range of micro-eV, corresponding to a range of $10^{12}$ GeV $  <  f_a  < 10^{14}$ GeV, favoring the highest helicity spin-flip cross-sectional values
presented in this work.\\


\section{Concluding remarks and outlook}
\label{s7}

In this report, the cross section for neutrino helicity spin-flip obtained
from a new $f(R,T)$ model of gravitation with dynamic torsion field introduced
by one of the authors in~\cite{Cirilo-Lombardo:2013cua}, was phenomenologically analyzed. To this
end, due the logarithmical energy dependence of the cross section, the relation
with the axion decay constant $f_{a}$ (Peccei-Quinn parameter) was used.
Consequently, the link with the phenomenological energy/mass window is found
from the astrophysical and high energy viewpoints. The important point is that,
in relation with the torsion vector interaction Lagrangian, the $f_{a}$
parameter gives an estimate of the torsion field strength that can variate with
time within cosmological scenarios~\cite{CiriloLombardo:2010zza,CiriloLombardo:2011zz}, potentially capable of modifying the overall leptogenesis picture.

\section{Acknowledgements}

D.J. Cirilo-Lombardo is grateful to the Bogoliubov Laboratory of Theoretical
Physics-JINR and CONICET-ARGENTINA for financial support and to Andrej B.
Arbuzov and Peter Minkowski for discussion and insights. Jilberto Zamora-Saa was
supported by Fellowship Grant \textit{Becas Chile} N$^{o}$74160012, CONICYT.

\section*{References}

\bibliography{mybibfile}

\end{document}

%% file: def_JZS.tex
\usepackage{amsmath,amsfonts,amssymb,amsthm,bm,bbm,bbold,braket,color,dsfont,fancybox,hyperref,
srcltx,subfigure,slashed,ucs,yfonts,cancel,xfrac,verbatim,xcolor}
\usepackage{floatrow}
\usepackage{graphicx}
\usepackage[outercaption]{sidecap} 
\usepackage[inline,shortlabels]{enumitem}
\usepackage{nicefrac}
\usepackage{multirow}
\DeclareMathAlphabet{\mathpzc}{OT1}{pzc}{m}{it}
\definecolor{gold}{rgb}{1,0.8,0}
\definecolor{nara}{rgb}{1,0.4,0.1}
\definecolor{goldo}{rgb}{1,0.7,0}
\definecolor{greeno}{rgb}{0,0.8,0}
\def\bes{\begin{subequations}}
\def\ees{\end{subequations}}
\def\be{\begin{equation}}
\def\ee{\end{equation}}
\def\bea{\begin{eqnarray}}
\def\eea{\end{eqnarray}}
\def\ba{\begin{eqnarray}}
\def\ea{\end{eqnarray}}
\def\bear{\begin{array}}
\def\eear{\end{array}}
\newcommand{\bpm}{\begin{pmatrix}}
\newcommand{\epm}{\end{pmatrix}}
\newcommand{\BM}{\left(\begin{array}}		
\newcommand{\BMC}{\left[\begin{array}}		

\newcommand{\EM}{\end{array}\right)}		
\newcommand{\EMC}{\end{array}\right]}		

	% ...

\newcommand{\com}[1]{}